\documentclass{PoS}
\usepackage{amsmath}

\title{The leading disconnected contribution to the anomalous magnetic moment of
the muon}

\ShortTitle{The leading disconnected contribution to the anomalous magnetic
moment of the muon}

\author{Anthony~Francis$^{1,2}$,
\speaker{Vera~Gülpers}$^{1,2}$, Benjamin~Jäger$^3$, Harvey~Meyer$^{1,2}$,
Georg~von~Hippel$^{1}$,
Hartmut~Wittig$^{1,2}$\\%
        $^{1}$PRISMA Cluster of Excellence, Institut für Kernphysik, Johannes
Gutenberg Universität Mainz, 55099 Mainz, Germany\\
	$^{2}$Helmholtz Institute Mainz, Johannes Gutenberg Universität Mainz,
55099 Mainz, Germany \\
	$^3$Department of Physics, College of Science, Swansea
University, SA2 8PP Swansea, UK\\
        E-mail: \email{guelpers@kph.uni-mainz.de}}

\abstract{The hadronic vacuum polarization can be determined from the vector
correlator in a mixed time-momentum representation. We explicitly calculate the
disconnected contribution to the vector correlator, both in the $N_f = 2$ theory
and with an additional quenched strange quark, using non-perturbatively
$O(a)$-improved Wilson fermions. All-to-all propagators are computed using
stochastic sources and a generalized hopping parameter expansion. Combining the
result with the dominant connected contribution, we are able to estimate an
upper bound for the systematic error that arises from neglecting the
disconnected contribution in the determination of $(g-2)_\mu$.}

\FullConference{The 32nd International Symposium on Lattice Field Theory,\\
		23-28 June, 2014\\
		Columbia University New York, NY}

\begin{document}

\section{Introduction}
\begin{figure}
\centering
 \includegraphics{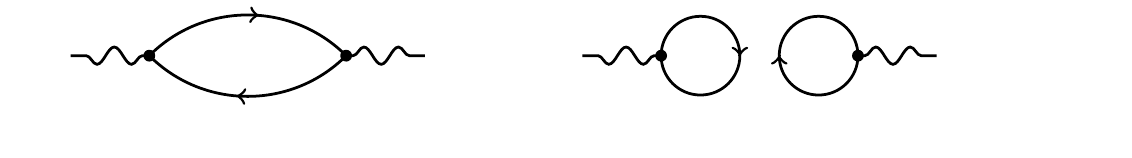}
 \caption{The connected and the disconnected contribution to the hadronic
vacuum polarization.}
\label{fig:diagrams}
\end{figure}
The anomalous magnetic moment of the muon $a_\mu$ is one of the most precisely
measured quantities in particle physics. A deviation of $\approx 3\sigma$
between the experimental and the theoretical value has persisted for many years.
From the theory side, the largest fraction of the error
comes from the hadronic vacuum contribution (hvp), which is the leading order
QCD contribution to $a_\mu$. Currently, the best estimate of the hvp relies on
a semi-phenomenological approach using the cross section
of $e^+\,e^-\,\rightarrow$ hadrons. In the
past few years, a lot of effort has been undertaken to calculate the hvp from
first principles using lattice techniques
\cite{amuaubinblum,amuETMC,amuBoyle,amuMainz}. However, the quark-disconnected
contribution to the hvp is generally neglected. This may be a significant
source of systematic error, since in partially quenched chiral perturbation
theory, it was estimated that the disconnected contribution could be as large as
$-10\%$ of the connected one \cite{chipt}.\par
We explicitly compute the disconnected contribution to the hvp with
$\mathcal{O}(a)$-improved Wilson fermions using the mixed-representation method
\cite{mixedrep,mixedrepetmc}, where the hadronic vacuum polarization is
calculated using the vector correlator
\begin{equation}
   G^{\gamma\gamma}(x_0) = -\frac{1}{3}\,\int\!\textnormal{d}^3x
\left<j^\gamma_k(x)j^\gamma_k(0)\right>\hspace{0.5cm}\textnormal{with}\hspace{
0.3cm } j^\gamma_k = \frac{2}{3} \overline{u}\gamma_k u - \frac{1}{3}
\overline{d}\gamma_k d + \ldots
\end{equation}
as follows:
\begin{equation}
 \hat{\Pi}(Q^2) = 4\pi^2\int\limits_0^\infty
\textnormal{d}x_0\,G^{\gamma\gamma}(x_0)\left[x_0^2
-\frac{4}{Q^2} \sin^2\left(\frac{1}{2}Qx_0\right)\right]\,.
\label{eq:mixedrepmethod}
\end{equation}
The vector correlator $ G^{\gamma\gamma}(x_0)$ receives a connected
and a disconnected contribution as shown in figure \ref{fig:diagrams}.
We calculate the required disconnected quark loops using
stochastic sources and a hopping parameter expansion as described in
\cite{mypaper}.

\section{Results for the vector correlator}
\begin{figure}[h]
 \centering
      \includegraphics[width=0.48\textwidth]{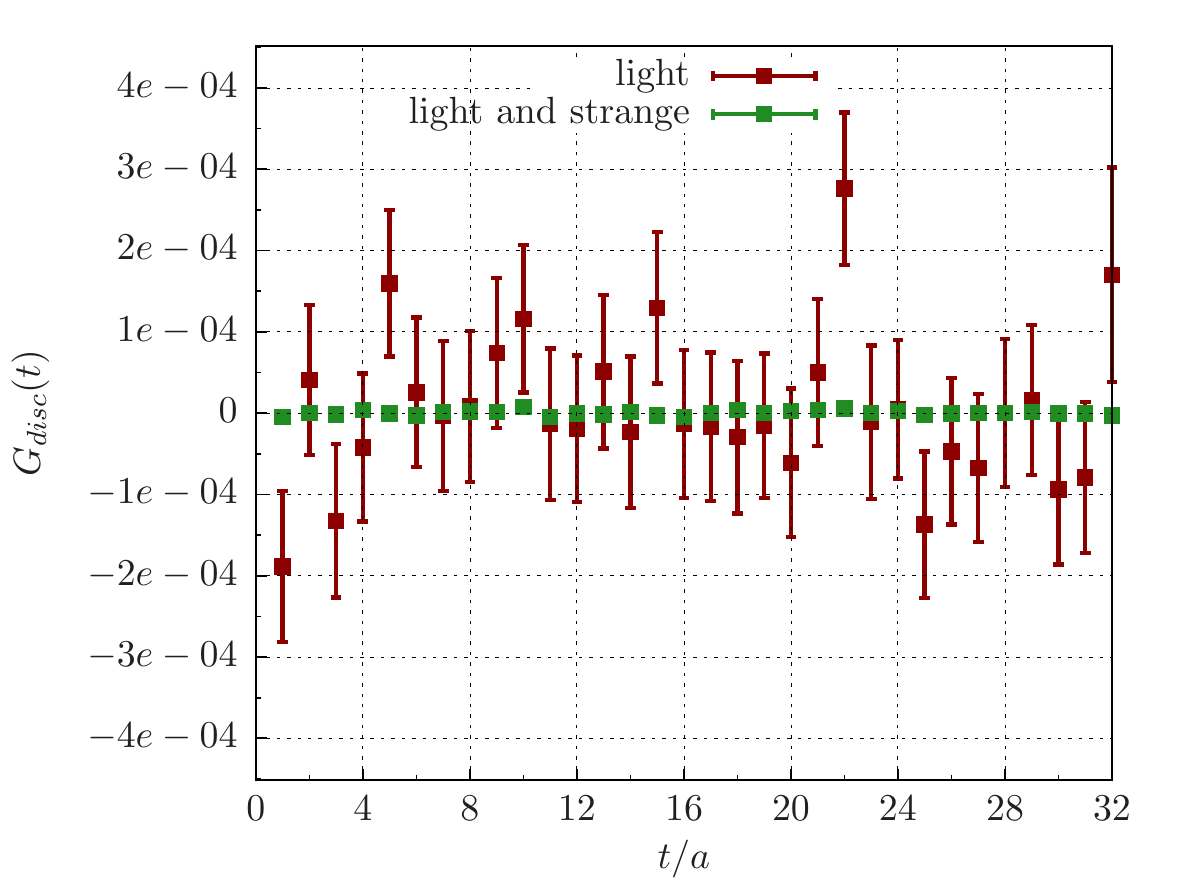}
   \includegraphics[width=0.48\textwidth]{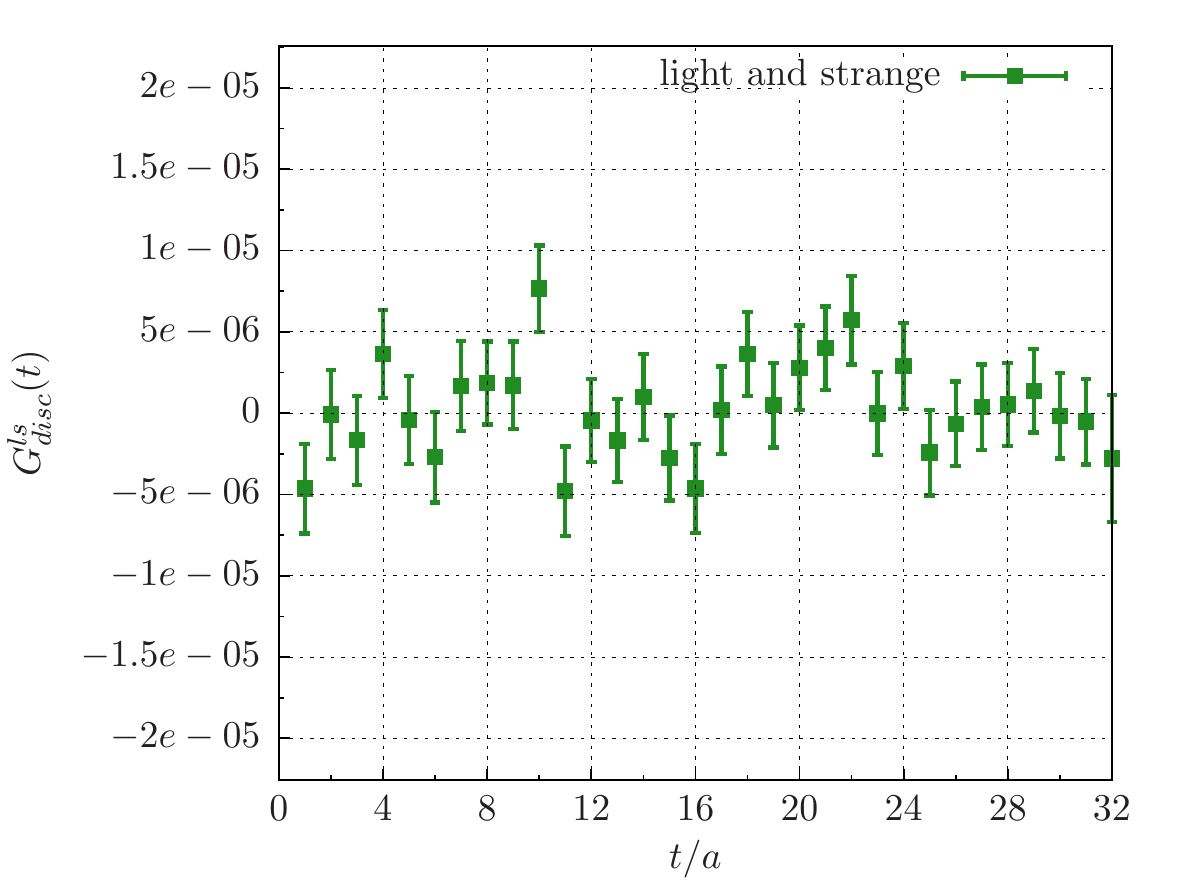}
 \caption{The disconnected vector correlator for light quarks (red) and
combined light and strange quarks (green). Note, that the scales on both plots
are different.}
\label{fig:disccorr}
\end{figure}

In the following we will concentrate on the vector correlator for light and
strange quarks combined. The corresponding electromagnetic current
\begin{equation}
	  j^{\ell s}_\mu= j^{\ell}_\mu +j^{s}_\mu=
\underbrace{\frac{1}{2}\,(\overline{u}\gamma_\mu u -
	      \overline{d}\gamma_\mu d)}_{I=1,\,\,\,\,j^\rho_\mu} +
\underbrace{\frac{1}{6}	\,(\overline{u}\gamma_\mu u +\overline{d}\gamma_\mu
d-2\overline{s}\gamma_\mu s)}_{I=0}
\end{equation}
can be split into an isovector part corresponding to the $\rho$-current and an
isoscalar part. Performing the Wick contractions one finds for the light and
strange vector current
\begin{equation}
 G^{\ell s}(t)=\frac{5}{9} G^{\ell }_\textnormal{con}(t) +
  \frac{1}{9}G^{s}_\textnormal{con}(t) + \frac{1}{9}
 G^{\ell s}_\textnormal{disc}(t)\,\hspace{1cm}\textnormal{with}\hspace{0.5cm}
G^{\ell }_\textnormal{con}(t) = 2G^{\rho\rho}(t)
\label{eq:corr}
\end{equation}
For convenience, we consider the disconnected correlator $G^{\ell
s}_\textnormal{disc}(t)$ for light and strange quarks combined, since one can
write the disconnected Wick contractions as
\begin{equation}
\begin{aligned}
 G^{\ell s}_\textnormal{disc}(x_0) &= -\!\!\int\!\textnormal{d}^3x
 \left<j^{\ell s}_k(x)j^{\ell s}_k(0)\right>_\textnormal{disc}\\
&=-\!\!\int\!\textnormal{d}^3x
 \left<(j^{\ell }_k(x) -
j^{s}_k(x))\,(j^{\ell }_k(0)-j^{s}_k(0))\right>_\textnormal{disc}\,,
\end{aligned}
\label{eq:disccorr}
\end{equation}
i.e.\ we only need differences of light and strange quark loops. Thus, we
expect that stochastic noise can be canceled when light and strange quark loops
are calculated using the same stochastic sources. Figure \ref{fig:disccorr}
shows our results for the disconnected correlator for light quarks only in red
and for combined light and strange quarks in green for the E5 ensemble (cf.
table \ref{tab:ensembles}). As expected, we find that the stochastic error for
the combined light and strange disconnected correlator is significantly smaller
than the error on the light quark correlator alone.
Although we can reduce the statistical error significantly when light and
strange loops are calculated with the same stochastic sources, we find that
the disconnected correlator $ G^{\ell s}_\textnormal{disc}(x_0)$ is still
consistent with zero within our current accuracy.
\par
\begin{figure}[h]
 \centering
      \includegraphics[width=0.48\textwidth]{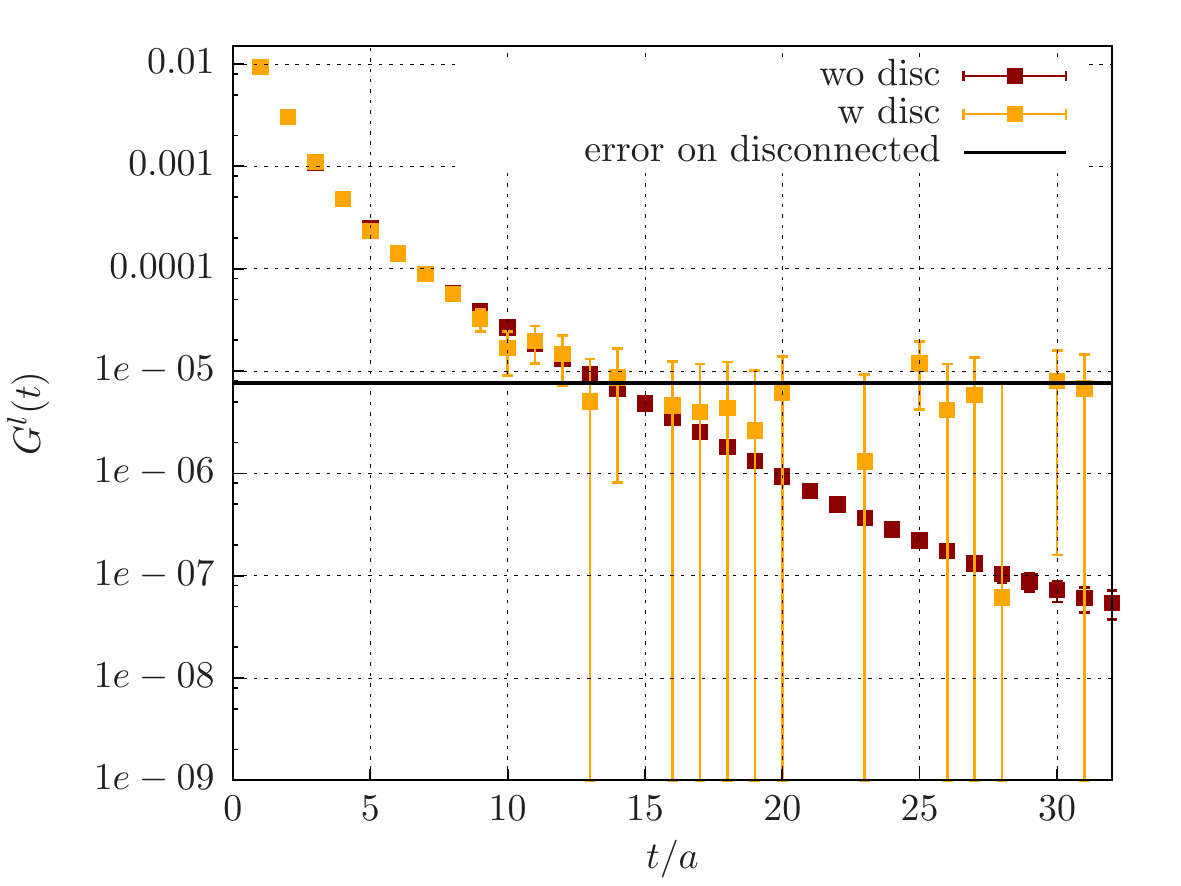}
   \includegraphics[width=0.48\textwidth]{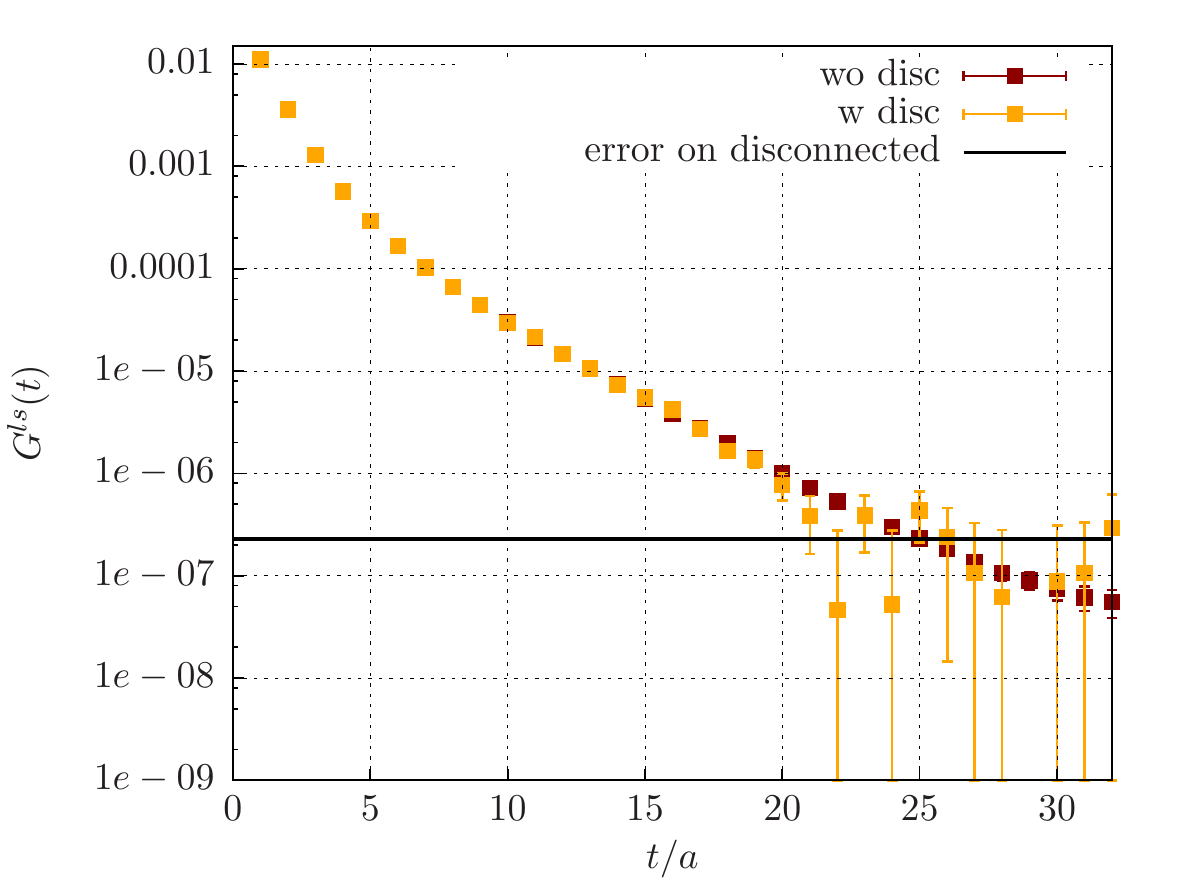}
 \caption{The connected (red) and the total (yellow) vector correlator for
light quarks (left) and light and strange quarks (right). The horizontal line in
both plots shows level of the statistical error on the disconnected
contribution.}
\label{fig:vectorcorr}
\end{figure}

We can add the disconnected correlator to the connected one to
obtain the total vector correlator. Figure \ref{fig:vectorcorr} shows the
connected (red) and the total vector correlator (yellow) for the E5 ensemble.
Results for light quarks as well as light and strange quarks combined are shown
on the left- and the right-hand side, respectively. The horizontal line in both
plots shows the level of the statistical error on the disconnected contribution,
i.e.\ it indicates the point from which on our total vector correlator is
dominated by the noise of the disconnected contribution. This point sets in for
significantly larger euclidean times in the case of the combined light and
strange quark correlator.
\par
Although we do not find a non-vanishing signal for the disconnected correlator,
we can still use our results to give a limit for the maximum possible
contribution to the hadronic vacuum polarization from quark-disconnected
diagrams. Here, we will solely consider the case of combined light and strange
quarks, for which the statistical error is significantly smaller.

\section{The vector correlator for large euclidean times}
\begin{figure}[h]
 \centering
      \includegraphics[width=0.46\textwidth]{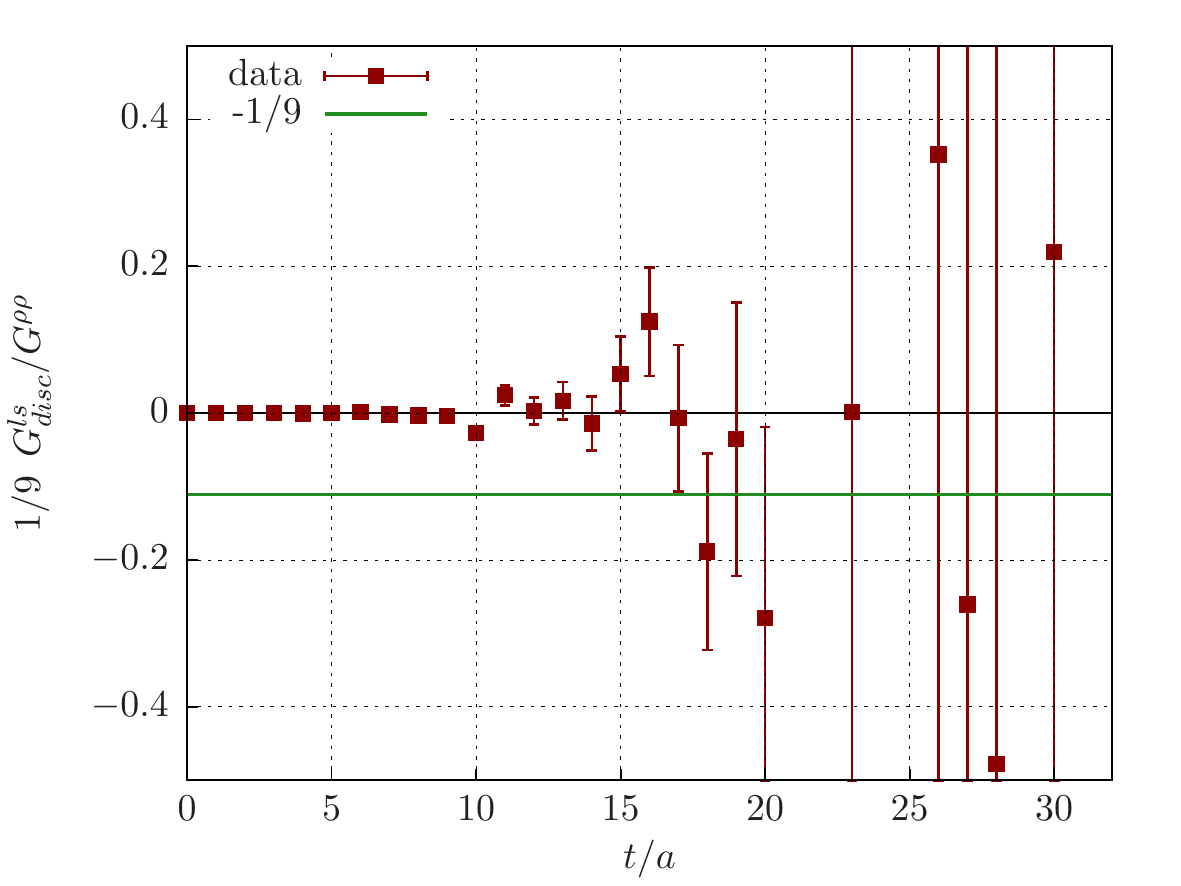}
   \includegraphics[width=0.46\textwidth]{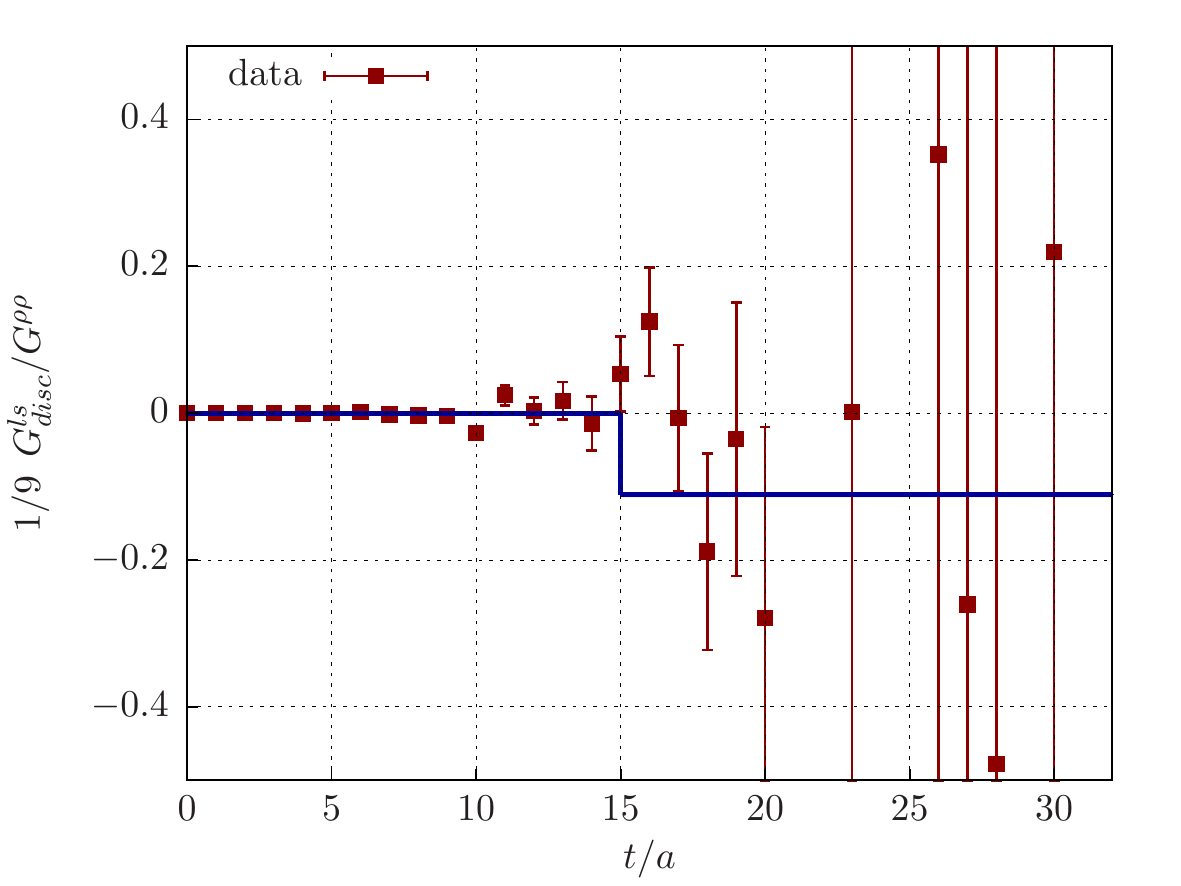}
 \caption{The ratio of the disconnected correlator and the $\rho$-correlator.
The green line on the left-hand side shows the asymptotic value. The blue line
on the right-hand side shows our conservative estimate for the disconnected
correlator.}
\label{fig:ratio}
\end{figure}

In order to estimate the maximum possible contribution from quark-disconnected
diagrams we require information about the behavior of the vector correlator for
large euclidean times in addition to our data.
For large euclidean times, the vector correlator is dominated by the isovector
part \cite{mixedrep}, due to its lower threshold:
\begin{equation}
 G^{\gamma\gamma}(t) =
G^{\rho\rho}(t)\left(1 + \mathcal{O}(e^{-m_\pi t})\right)\,.
\end{equation}
If we rewrite equation \eqref{eq:corr} as
\begin{equation}
{\frac{1}{9}\frac{G^{\ell s}_\textnormal{disc}(t)}{G^{\rho\rho}(t)} =}
\underbrace{\color{black}\frac{G^{\gamma\gamma}(t)-G^{\rho\rho}(t)}{G^{\rho\rho}
(t) } } _ { \rightarrow\,0\quad \text{for}\,\,t\rightarrow\infty}
{\color{black}-
\frac{1}{9}}\underbrace{ \color{black}\left(1 +
2\frac{G^{s}_\textnormal{con}(t)}{G^{\ell }_\textnormal{con}(t)}\right)}_{
\rightarrow\,1\quad \text{for}\,\,t\rightarrow\infty}
\hspace{0.5cm}\longrightarrow\,\,\,-\frac{1}{9}\,,
\label{eq:ratio}
\end{equation}
we find an asymptotic value of $-1/9$ for the ratio of the light and strange
disconnected correlator $G^{\ell s}_\textnormal{disc}(t)$ to the
$\rho$-correlator for large euclidean times. This %
ratio \eqref{eq:ratio} is plotted against $t$ in figure~\ref{fig:ratio}. The
green line on the left-hand side shows the asymptotic value $-1/9$. As one can
see, we can clearly distinguish the ratio from its asymptotic value up to
$t\approx15a$. \par
To give a conservative upper limit for the disconnected contribution, we
assume that the ratio \eqref{eq:ratio} falls monotonically from zero to $-1/9$
at some point. Furthermore, our estimate for $G^{\ell s}_\textnormal{disc}(t)$
has to be consistent with both our data and with its theoretical asymptotic
value. 
Thus, the disconnected contribution would be maximized if the the ratio
were basically zero up to $t\approx 15a$ and then suddenly dropped to
$-1/9$, as indicated by the blue line. 
If we take this as an estimate of the
disconnected vector correlator, we can give a conservative upper bound for the
magnitude of the disconnected contribution to $a_\mu$.

\section{Hadronic vacuum polarization and $a_\mu$}
From the vector correlator, one can calculate the hadronic vacuum polarization
(cf. equation~\eqref{eq:mixedrepmethod}).
We calculate $\hat{\Pi}^{\ell s}(Q^2)$ once only for the connected vector
correlator (for the details of the analysis, see \cite{anthonyproc}) and once
with the disconnected estimate as described above, i.e.\
\begin{itemize}
\item for $t\leq15a\approx1$~fm, the vector correlator is well described by
the connected part, i.e. we use
\begin{equation}
 G^{\ell s}(t) = \frac{5}{9} G^{\ell }_\textnormal{con}(t) +
\frac{1}{9}G^{s}_\textnormal{con}(t)
\end{equation}
\item for $t>15a$ we use the asymptotic value 
$\frac{1}{9}\,\,G^{\ell s}_\textnormal{disc}(t)/G^{\rho\rho}(t)
= -1/9$ as an upper bound for the disconnected part,
\begin{equation}
 G^{\ell s}(t) = \frac{5}{9} G^{\ell }_\textnormal{con}(t) +
\frac{1}{9}G^{s}_\textnormal{con}(t)-\frac{1}{9}G^{\rho\rho}(t)\,.
\end{equation}
\end{itemize}
\begin{figure}[h]
 \centering
\includegraphics[width=0.46\textwidth]
{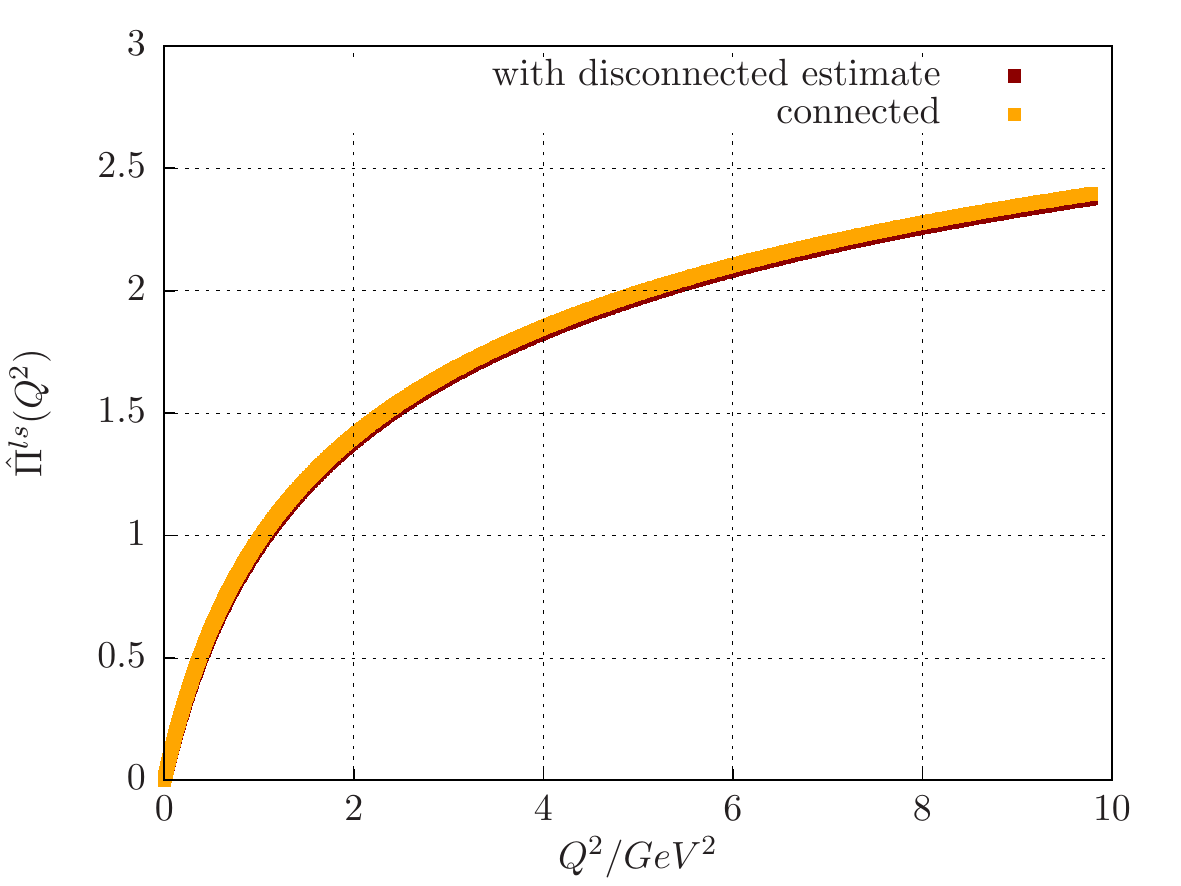}
\includegraphics[width=0.46\textwidth]{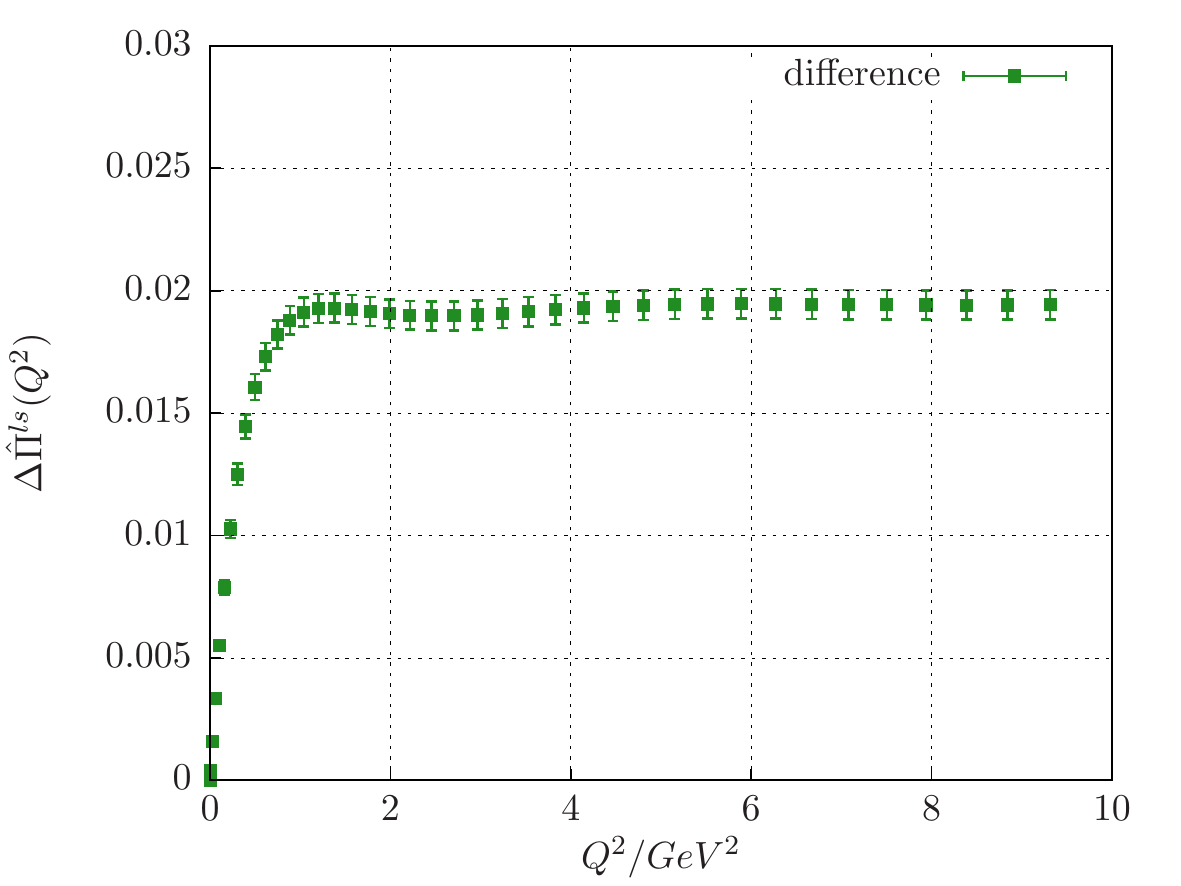}
 \caption{The plot on the left-hand side shows the vacuum polarization from the
connected correlator (yellow) and for the correlator with an estimate for the
disconnected contribution. The plot on the right-hand side shows their
difference.}
\label{fig:vacpol}
\end{figure}
The left-hand side of figure \ref{fig:vacpol} shows the vacuum contribution for
both cases. As expected, the vacuum polarization with the disconnected estimate
is smaller than the vacuum polarization from the connected contribution only,
since for large euclidean times $G^{\ell s}_\textnormal{disc}(t)$ has the
opposite sign than the connected correlator. Since the difference between the
two curves is small, the right hand side of figure \ref{fig:vacpol} shows their
difference, which is larger than the statistical error on $\hat{\Pi}^{\ell
s}(Q^2)$.
\par
From the vacuum polarization, one can now calculate the hadronic contribution
to the anomalous magnetic moment of the muon \cite{kernel,blum},
\begin{equation}
 a_\mu^{\textnormal{hvp}} = \left(\frac{\alpha}{\pi}\right)^2
\int\limits_0^\infty \textnormal{d} Q^2\,\frac{1}{Q^2} K(Q^2)\,\hat{\Pi}(Q^2)\,,
\end{equation}
with an electromagnetic kernel function $K(Q^2)$. We calculate
$a_\mu^{\textnormal{hvp}}$ once for the vacuum polarization for the connected
part only, and once for the vacuum polarization which includes the disconnected
estimate. For the E5 ensemble, we find that with the disconnected estimate the
result for $a_\mu^{\textnormal{hvp}}$ is $\approx 3.5\%$ smaller. One has to
keep in mind that this is a conservative upper limit, and that the disconnected
contribution to $a_\mu^{\textnormal{hvp}}$ could also be much smaller. We use
the $3.5\%$ as an upper bound for a systematic error that arises when the
disconnected contribution is neglected.
\par
\begin{table}[h]
 \centering
  \begin{tabular}{|cccccccc|}
\hline
$\beta$ &  $a [\textnormal{fm}]$ & lattice & $m_\pi [\textnormal{MeV}]$ &
$m_\pi L$ & Label & $N_{\textnormal{cnfg}}$ & $t_\textnormal{cut}$\\
\hline\hline
$5.3$ & $0.063$  & $64\times32^3$ & $451$ & $4.7$ &  E5 & $1000$ & $15$\\
$5.3$ & $0.063$  & $96\times48^3$ & $324$ & $5.0$ &  F6 & $300$ & $13$\\
$5.3$ & $0.063$  & $96\times48^3$ & $277$ & $4.3$ &  F7 & $250$ & $13$\\
\hline
 \end{tabular}
\caption{The CLS ensembles used for the calculation of the disconnected
contribution to the hadronic vacuum polarization.}
\label{tab:ensembles}
\end{table}
So far, we have done this calculation for three different gauge ensembles,
which are listed in table \ref{tab:ensembles}. For the ensembles F6 and F7 we
have less statistics than for E5, and we can not resolve the ratio of
disconnected correlator and $\rho$-correlator as well as for E5. Thus, we
choose a slightly smaller value $t_\textnormal{cut}$ up to which we neglect the
disconnected correlator and from which on we use the asymptotic value. For both
ensembles we find a upper limit for the disconnected contribution of
$\approx5\%$.
\par
\begin{figure}[h]
\centering
 \includegraphics[width=0.55\textwidth]{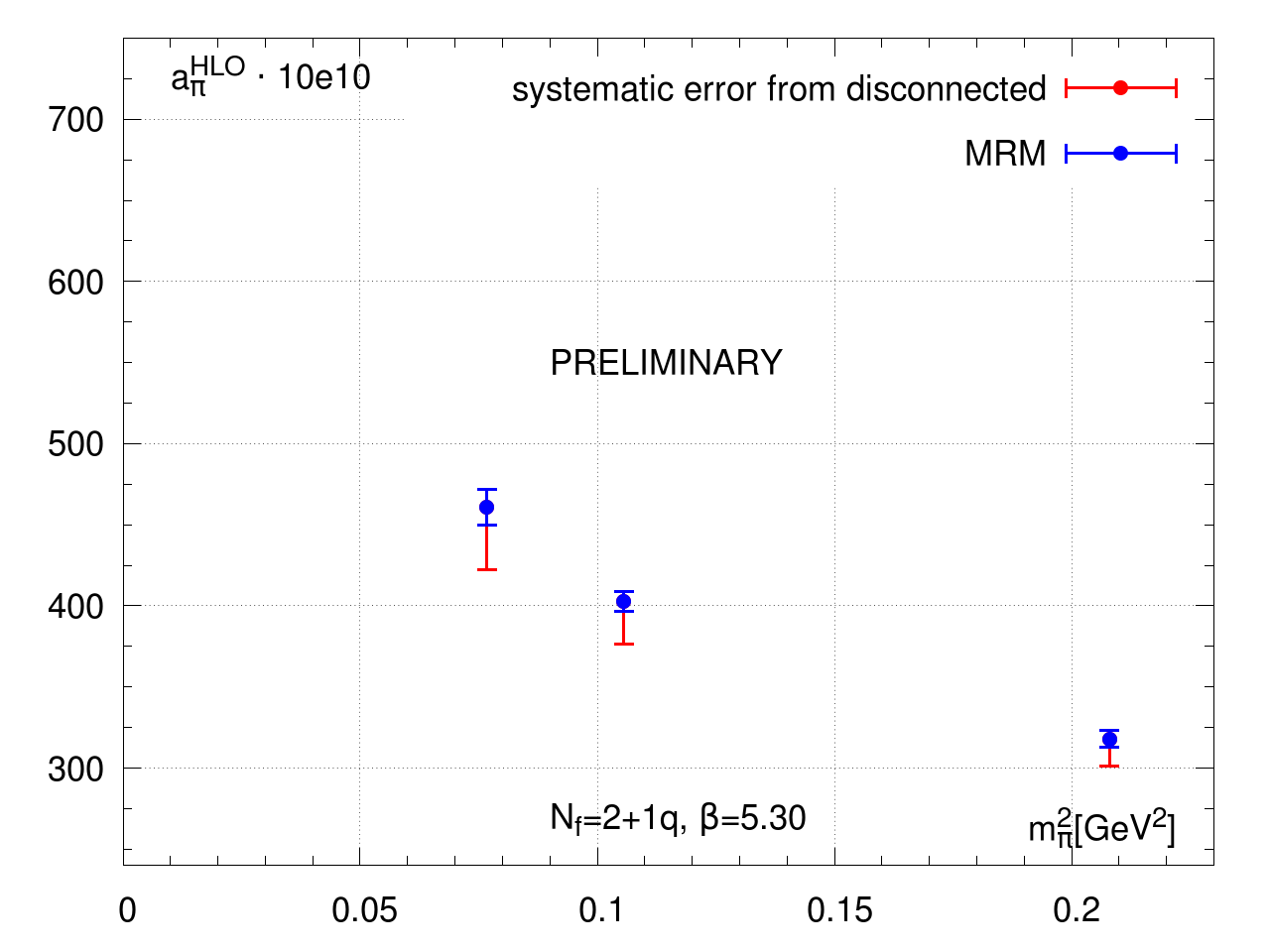}
\caption{$a_\mu^{\textnormal{hvp}}$ plotted against the pion mass. Blue points
show the results from the connected correlator. The red error bars show the
maximum systematic error from neglecting the disconnected contribution.}
\label{fig:amu}
\end{figure}
Figure \ref{fig:amu} shows the results for $a_\mu^{\textnormal{hvp}}$ plotted
against $m_\pi^2$. The blue points show the results for the connected
correlator and are the same as in \cite{anthonyproc}. The red error bars
denote the maximum systematic error from neglecting the disconnected
contribution. One can see that this systematic error is larger than the
statistical error on $a_\mu^{\textnormal{hvp}}$. Thus, improving the accuracy
of the QCD prediction of the hadronic contribution to $a_\mu$ requires
improvements to the computation of the disconnected contribution.

\section{Conclusions}
We have explicitly calculated the disconnected vector correlator for light and
strange quarks. Since the disconnected correlator depends only on the
difference of light and strange propagators, the statistical error can be
significantly reduced when using the same stochastic sources for light and
strange loops. However, we still find that the disconnected vector correlator
is consistent with zero within our current accuracy. Using the asymptotic
behavior of the vector correlator for large euclidean times, we are able to
give an upper limit for the disconnected contribution to
$a^{\textnormal{hvp}}_\mu$. The lattice data shows that up to some time
$t_\textnormal{cut}$ the vector correlator is well described by the connected
contribution only, and thus the disconnected one can be neglected. From this
time on, we use the asymptotic value for $G^{\ell s}_\textnormal{disc}(t)$.
This allows us to give an upper estimate for the systematic error from
neglecting the disconnected contribution, which we find to be of the order of
$4-5\%$. We note however, that this estimate is as conservative as possible, and
that the disconnected contribution might be much smaller. Nevertheless, any
further reduction of the error on the calculation of the hadronic contribution
to the anomalous magnetic moment of the muon from lattice QCD requires an
improvement of the computation of the disconnected contribution.

\par
\vspace{0.4cm}
\textbf{Acknowledgements}
{\footnotesize  Our calculations were performed on the  dedicated QCD platforms
``Wilson'' at the Institute for Nuclear Physics, University of
Mainz, and ``Clover'' at the Helmholtz-Institut Mainz. We thank Dalibor
Djukanovic and Christian Seiwerth for technical support.
We are grateful for computer time allocated to project HMZ21 on the BlueGene
computer \mbox{``JUQUEEN''} at NIC, J\"ulich. This research has been supported
in part by the DFG in the SFB~1044. We are grateful to our colleagues in the CLS
initiative for sharing ensembles.
}

\end{document}